\documentclass[12pt]{iopart}

\begin{document}

\title{Effective diffusion constant in a two dimensional medium of
charged point scatterers}

\author{D.S. Dean\dag\ddag\ I.T. Drummond\dag and R.R. Horgan\dag }

\address{\dag \ DAMTP, CMS, University of Cambridge, Cambridge, CB3
0WA, UK}

\address{\ddag\ IRSAMC, Laboratoire de Physique Quantique,
Universit\'e Paul Sabatier, 118 route de Narbonne, 31062 Toulouse
Cedex 04, France}

\begin{abstract}
We obtain exact results for the effective diffusion constant of a two
dimensional Langevin tracer particle in the force field generated by
charged point scatterers with quenched positions.  We show that if the
point scatterers have a screened Coulomb (Yukawa) potential and are
uniformly and independently distributed then the effective diffusion
constant obeys the Volgel-Fulcher-Tammann law where it vanishes. Exact
results are also obtained for pure Coulomb scatterers frozen in an
equilibrium configuration of the same temperature as that of the
tracer.

\end{abstract}



\section{Introduction}
The bulk transport properties of random media are of great importance
in physics and engineering. The computation of the bulk diffusivity,
conductivity, permeability and dielectric constant from the
statistical properties of their local fluctuations is referred to as
homogenisation. These bulk quantities describe the transport
properties over large length and time scales and will normally have
well defined values when the statistics of the local fluctuations are
homogeneous under translation. We will consider situations that are
also statistically isotropic since a breakdown of isotropy involves
further levels of complexity, important and interesting though they
are.

In this paper we compute the effective diffusion constant for tracer
particles travelling through a medium of fixed but randomly
distributed centres of force. We we assume for simplicity that the
centres of force are all identical but may have either positive or
negative charge with equal probability. The tracer therefore
experiences a randomly fluctuating potential with zero mean. This
model of a disordered medium is a reasonable, if simplified,
description of many real physical systems where the random potential
is induced by impurities inserted in a regular background.

Many studies of diffusion in a random potential have been based on the
assumption that it can be represented by a Gaussian field
\cite{ddh1,dc,ddh2,ddh3,ddh4}. Diffusion in a non-Gaussian field has
been studied using perturbation theory \cite{dhs}. However the
calculation is significantly more complicated than that for the
Gaussian case. The significance of our investigation is that we obtain
exact results for a class of problems, diffusion in random potentials
of point scatterers, that are non-Gaussian and for which few results
are known. The results are exact in one and two-dimensions but
experience suggests that they may be indicative for higher dimensional
cases.

\section{The Model}
The position $X_i$, of a particle, of negligible inertia, subject to a
white noise, $w_i(t)$, and a potential $\phi(x)$, satisfies the
equation
\begin{equation}
{\dot X}_i=w_i(t)+\lambda_0\nabla_i\phi(X)~~,
\end{equation}
where
\begin{equation}
\langle w_i(t)w_j(t')\rangle=2\kappa_0\delta_{ij}\delta(t-t')~~,
\end{equation}
and $\langle\cdots\rangle$ denotes an average over the white noise.  
The Einstein relation implies that the local, or bare, diffusivity
$\kappa_0$ and the coupling to the potential gradient $\lambda_0$ are
related by the equation
\begin{equation}
\frac{\lambda_0}{\kappa_0}=\beta=\frac{1}{T}~~,
\end{equation}
where $T$ is the absolute temperature in appropriate units.  The
probability density, $p(x,t)$, for the position of the particle obeys
\begin{equation}
{\partial p(x,t)\over \partial t} = \kappa_0\nabla^2 p +
\lambda_0\nabla \cdot (p \nabla \phi)~~.
\label{deg} 
\end{equation}
The effective diffusivity, $\kappa^{(g)}_e$, appropriate to the
dispersion of the particle at large times and distances is
\begin{equation}
\kappa^{(g)}_e=\frac{1}{2D}\lim_{t\rightarrow\infty}\frac{\langle
X_t^2\rangle}{t}~~,
\end{equation}
where $D$ is the dimension of space. The mean squared displacement
of the particle is given by
\begin{equation}
\langle X_t^2\rangle=\int d^Dx\ x^2 p(x,t)~~.
\end{equation}
The superscript $g$ is used to denote the fact that we are considering
the effective diffusion constant for a particle diffusing in a {\em gradient} 
field.

The description of the model is completed by specifying the structure
of the potential $\phi(x)$ in terms of the scattering centres and their
charges. In this paper we will consider potentials of the type
\begin{equation}
\phi(x)=\sum_{n=1}^{N}q_nV(x-x_n)~~,
\end{equation}
generated by $N$ scatterers frozen in the volume $V$.
Here $q_n=\pm 1$ is the charge on the scatterer $n$, $x_n$ is
its position and $V(x)$ the potential at the point $x$
due to a positively charged background particle at the origin.
In this paper we shall consider distributions of the $x_n$ and 
$q_n$ such that the field $\phi$ is statistically invariant
under global transformation $\phi\to -\phi$. This is clearly the case
for distributions of the $x_n$ which are homogeneous and isotropic
along with the condition that the distribution of the $q_n$
is invariant under the global charge transformation $q_n \to -q_n$.
An example of this type of distribution is one where the positions
$x_n$ are distributed uniformly and independently in the volume $V$
and the charges are taken to be independent and $\pm 1$ with probability 
${1\over2}$. In this case the disorder average is given by
\begin{equation}
\langle {\cal O} \rangle_d =\prod_{n}\sum_{q_n=\pm 1}
\frac{1}{2V}\int d^Dx_n~~{\cal O}~~. \label{uni1}
\end{equation}
One could also consider an strictly
electroneutral ensemble where $N/2$ of the scatterers
have the charge $\pm 1$, here the disorder average is thus
\begin{equation}
\langle {\cal O} \rangle_d =\prod_{n}{1\over V}
\int d^D x_n~~{\cal O}~~, \label{uni2}
\end{equation}
we shall see however in the thermodynamic limit 
these two problems have the same diffusion constant.
Another realisation of an ensemble of the type mentioned above is
one where a system of  $N/2$ positive and negative charges, interacting via the
Hamiltonian
\begin{equation}
H =  \sum_{i<j}^N q_i q_j V(x_i-x_j)~~,
\end{equation}
are allowed to interact and equilibriate at some inverse temperature 
$\beta'$, they are then frozen in this equilibrium configuration
giving a distribution of the $x_n$ of 
\begin{equation}
P(\{x_i\}) = {1\over Z(\beta')}\exp\left(-\beta'\sum_{i<j}^N q_i q_j V(x_i-x_j)
\right)~~,\label{eq}
\end{equation}
where $Z(\beta')$ is the partition function for the system. Clearly 
the uniform distribution of equation (\ref{uni2}) is recovered in the limit 
$\beta'\to 0$.

\section{Associated permeability model}
In a previous paper we showed that there is a strong connection
between our gradient flow model and the problem of computing the
effective long range diffusivity, $\kappa_e^{(p)}$, of particle with a
locally random diffusivity field $\kappa(x)$~. 
A Langevin particle diffusing in a random medium with local
diffusivity $\kappa(x)$ has a probability density which satisfies
\begin{equation}
{\partial p(x,t)\over \partial t} = \nabla \cdot \kappa(x)\nabla p~~.
\label{dep}
\end{equation}

In this case also the long time transport can also  be partially described
by the asymptotic behaviour of the mean squared displacement.
\begin{equation}
\langle X_t^2\rangle \simeq 2D\kappa_e^{(p)}t~~,
\end{equation}
where $\kappa_e^{(p)}$ is the bulk or effective diffusivity of the medium.

The problem of calculating $\kappa_e^{(p)}$ is also equivalent
calculating the bulk permeability of a porous medium, where 
the flow is described by Darcy's law \cite{mat}, given that the 
local permeability is  $\kappa(x)$. It is also equivalent
to calculating the effective conductivity or effective dielectric constant in
media where $\kappa(x)$ represents the local conductivity or dielectric
constant. The superscript $p$ here is thus used to denote
that the effective diffusion constant corresponds to an effective
{\em permeability} problem. The problem of calculating $\kappa_e^{(p)}$ 
has been considered by
numerous authors via exact relations in one and two dimensions
\cite{mat,kel,dy1,dy2} and via perturbation techniques in three
dimensions \cite{king,dh,dew,abin}.

The relevant connection between the permeability problem and the
gradient flow problem emerges when the local diffusivity or
permeability is chosen to be
\begin{equation}
\kappa(x) = \kappa_0\exp\left(\beta\phi(x)\right)~~. \label{eqkappa}
\end{equation}
In \cite{ddhd} it was shown that the bulk permeability or effective
diffusion constant for equation (\ref{dep}) with a local diffusivity given by
equation (\ref{eqkappa}) and that for the gradient flow problem
equation (\ref{deg}) with the same field $\phi$ are related by
\begin{equation}
\frac{\kappa_e^{(g)}}{\kappa_e^{(p)}}=\frac{\kappa_0}{\overline
\kappa}~~,
\label{ddhd}
\end{equation}
where
\begin{equation}
{\overline \kappa} = {1\over V} \int d^Dx\ \kappa(x)~~,
\end{equation}
The relation  equation (\ref{ddhd}) holds in all
dimensions where the corresponding $\kappa_e$ exist.
In a statistically homogeneous system  with short range
correlations in the field $\phi$, in the limit of large $V$ we expect
$\overline{\kappa}$ to be self averaging, or realisation independent, 
{\em i.e.}
\begin{equation}
{\overline \kappa} = 
\langle{\overline \kappa}\rangle_d =  \langle {1\over V}\int d^{D} x\ 
\kappa(x)\rangle_d~~,\label{sa}
\end{equation}

\section{Duality relation and the two dimensional result}

For two-dimensional systems we can obtain exact results by combining
equation (\ref{ddhd}) with a standard duality theorem \cite{kel,dy1,dy2}. 
For completeness we
derive the relevant result in a form appropriate to the problem under
consideration.

As a preliminary we note that when the fields $\phi(x)$ and $-\phi(x)$ are
statistically equivalent it follows that local diffusivities
$\kappa(x)$ and $\kappa'(x)=\kappa_0^2/\kappa(x)$~ are also
statistically equivalent. We view  equation (\ref{dep}) 
as describing the flow of the density $p(x)$ through the
random medium.  The associated current is $j(x)=\kappa(x)\nabla
p(x)$~. In the steady state we have
\begin{equation}
\nabla\cdot(\kappa(x)\nabla p(x))=0~~, \label{hh}
\end{equation}
and on averaging over the ensemble of samples we have, assuming a
constant mean current and density gradient,
\begin{equation}
\langle j(x)\rangle=\langle \kappa(x)\nabla p(x)\rangle
                =\kappa_e^{(p)}\langle\nabla p(x)\rangle~~.
\end{equation}
In two-dimensions equation (\ref{hh}) implies that there exists a dual field
$\chi(x)$ such that
\begin{equation}
j(x)=\kappa(x)\nabla p(x)=\kappa_0 n\times \nabla\chi(x)~~,
\end{equation}
where $n$ is a unit vector orthogonal to the two-dimensional plane of
the problem.  We also have the dual equation
\begin{equation}
\kappa'(x)\nabla\chi(x)=-\kappa_0n\times\nabla p(x)~~
\end{equation}
which implies
\begin{equation}
\nabla\cdot(\kappa'(x)\nabla\chi(x))=0~~.
\end{equation}
Since $\kappa'(x)$ is statistically equivalent to $\kappa(x)$ it
follows that
\begin{equation}
\langle\kappa'(x)\nabla\chi(x)\rangle=\kappa_e^{(p)}\langle\nabla\chi(x)\rangle~~,
\end{equation}
with the same effective permeability $\kappa_e^{(p)}$, as in the
original problem.  We have then the two dual results
\begin{equation}
\kappa_e^{(p)}\langle\nabla
p(x)\rangle=\kappa_0n\times\langle\nabla\chi(x)\rangle~~,
\end{equation}
and
\begin{equation}
\kappa_e^{(p)}\langle\nabla\chi(x)\rangle=-\kappa_0n\times\langle\nabla
p(x)\rangle~~.
\end{equation}
The consistency of these equations implies that
\begin{equation}
\kappa_e^{(p)}=\kappa_0~~. \label{2deq}
\end{equation}

Combining this result with equation (\ref{ddhd}) we obtain the general result
valid in two dimensions
\begin{equation}
\kappa_e^{(g)} =  {\kappa_0^2\over {\overline \kappa}}~~, \label{gen} 
\end{equation}
This equation (\ref{gen}) was previously used by the 
authors to solve the problem
of calculating $\kappa_e^{(g)}$ in a  homogeneous Gaussian field in two 
dimensions to give
\begin{equation}
\kappa_e^{(g)} = \kappa_0 \exp\left(-{\beta^2 \over 2} \Delta(0) \right)~~,
\label{gaus}
\end{equation}
where $\Delta$ is the connected two-point correlation function
defined  by
\begin{equation}
\Delta(x-x')=\langle\phi(x)\phi(x')\rangle_d -\langle\phi(x)\rangle_d 
\langle\phi(x')\rangle_d~~,
\end{equation}
the average again being over the sample disorder.
This 
result is particularly interesting as it showed that the renormalisation
group result for  $\kappa_e^{(g)}$ \cite{ddh1,dc}  in dimension $D$, given by 
\begin{equation}
\kappa_e^{(g)}(RG) = \kappa_0 \exp\left(-{\beta^2 \over D} \Delta(0) \right),
\label{eqrg}
\end{equation}
was exact in two dimensions. The equation (\ref{eqrg}) was already known to be 
exact in one dimension \cite{dc}.  
Indeed in one dimension the effective permeability
is given for a general homogenous $\kappa(x)$ by
\begin{equation}
\kappa_e^{(p)} = \left(\overline {1\over\kappa}\right)^{-1}
\end{equation} 
{\em i.e.} the harmonic mean of the local permeability. Using the result
equation (\ref{ddhd}) we thus obtain the solution of the corresponding 
gradient flow problem in one dimension to be 
\begin{equation} 
{\kappa_e^{(g)}\over \kappa_0} = 1/\left(\overline{\kappa} 
\overline{{1\over \kappa}}\right)
\end{equation}   

\section{Some Examples}

For the  problem of tracer particles moving through
scatterers uniformly and independently distributed in two dimensions,
we can combine equation (\ref{uni1}) with equation (\ref{sa}) to obtain
\begin{equation}
\fl {\overline \kappa} = \prod_n\left[\int {d^2x_n\over
V}\frac{1}{2}\sum_{q_n} \exp\left(\beta q_i V(-x_n)\right)\right] =
\left[ 1 + {1\over V} \int d^2x\ \left(\cosh\left(\beta V(x)\right)
-1\right) \right]^N~~.
\end{equation}
We now take the thermodynamic limit, that is $N\rightarrow\infty$ at
fixed $\rho=N/V$ and obtain the exact result
\begin{equation}
\kappa_e^{(g)} = \kappa_0\exp\left(-\rho\int d^2x\
\left(\cosh\left(\beta V(x)\right) -1\right)\right)~~.\label{equni}
\end{equation}
The distribution given by equation (\ref{uni2}) for an exactly neutral
system of scatterers of density $\rho$ also leads to the same result equation 
(\ref{equni}) in the thermodynamic limit.

It is interesting to review the relationship of this result to 
equation (\ref{gaus}) for the model in which $\phi(x)$ is a Gaussian field
\cite{ddh1,dc} with zero mean.  
In the case of uniformly distributed randomly charged scatterers the 
two-point correlation function of the field $\phi$ is given by 
\begin{equation}
\Delta(x-x')=\rho\int d^2x''\ V(x-x'')V(x'-x'')~~.
\end{equation}
Applying the  Gaussian ``prediction'' equation (\ref{gaus}) for 
the diffusivity  therefore gives
\begin{equation}
\kappa_e^{(g)}=\kappa_0\exp\left\{-\frac{\beta^2}{2}\rho\int d^2x\
V^2(x)\right\}~~.\label{gauss2}
\end{equation}
Comparing equations (\ref{gauss2}) and (\ref{equni}) we see that the
result for the Gaussian field is recovered from that of the point
scatterers in the limit $\beta \to 0$ while $\rho\beta^2 = c$ with $c$
a constant. This limit where the scatterer disorder acquires a
Gaussian character can also be found by examining the perturbation
theory for this problem and holds in all dimensions. It is interesting
to note that the Gaussian limit does not necessarily hold, as one may
have naively expected from the central limit theorem, simply in the
limit of large $\rho$.

It is illuminating to consider the particular case where the $V(x)$ is
a screened Coulomb potential (Yukawa interaction) in two
dimensions. We have
\begin{equation}
\nabla^2 V(x) - \mu^2 V(x) = - 2\pi \delta(x)
\end{equation}
where $\mu^{-1}$ is the screening length.  Writing $\vert x\vert = r$,
$V(x)=V(r)$ has the asymptotic behaviour
\begin{equation}
V(r) \propto\left\{\begin{array}{ll} -\ln(r)& r \ll \mu^{-1} \nonumber \\
                           \exp(-\mu r)/\sqrt{\mu r} & r \gg \mu^{-1}
                           \label{ylims}
\end{array}\right.
\end{equation}
From this we see that the integral
\begin{equation}
\int d^2x\ \left(\cosh\left(\beta V(x)\right) -1\right)
\end{equation} 
is convergent for large $r$ when $\mu >0$ and for small $r$ when
$\beta < 2$.  The importance of the screening is now revealed since
its removal, by setting $\mu=0$, leads to the integral becoming
infra-red divergent thus forcing the diffusion constant to vanish. It
would be interesting to investigate this regime separately since the
actual behaviour of the sample would presumably be sub-diffusive and
hence scale dependent.
 
The integral also diverges and the diffusivity goes to zero as $\beta$
approaches 2 from below, that is as the temperature cools to
$T=1/2$. It is possible to analyse qualitative nature of the
divergence since it is contained in the integral
\begin{equation}
\int_{r\le r_0} d^2x\ (\cosh(\beta V(x))-1)\simeq \int_0^{r_0}dr\
                                      Ar^{1-\beta} \simeq
                                      \frac{A}{2-\beta}~~,
\end{equation}
for some $A>0$ and sufficiently small $r_0$~. It follows that the
diffusivity behaves as
\begin{equation}
{\kappa_e^{(g)}\over \kappa_0} \approx B\exp\left( -{\rho A\over
2-\beta}\right)
\end{equation}
which is of the form of the Volgel-Fulcher-Tammann law reported in
fragile glass formers as they approach $T_g$. It is important to
recognise that the effect is one associated with {\it short}
distances. A qualitative light is shed on the circumstances by
considering the equilibrium tracer particle density in the presence of
the scattering centres. At any time only the nearest such scatterer
dominates the tracer particle probability density. If this has the
opposite sign of charge to the tracer particle then the probability
density is $h(x)$ where
\begin{equation}
h(x) = {\exp\left(\beta V(x)\right)\over \int_V d^Dx'\exp\left(\beta
V(x')\right)}~~,
\end{equation}
where we have placed the scatterer at the origin. Obviously this
exhibits the same divergence as before and the density only exists for
$\beta<2$~. The physical interpretation of this result is that the
tracer particle eventually becomes trapped at the point $x=0$ and thus 
diffusion is stopped.

We now consider a system where the configuration of the scatterers is
that obtained by freezing an equilibrium configuration of mobile
interacting scatterers at equilibrium at inverse temperature $\beta$,
the same inverse temperature as the tracer particle {\i.e.} with a distribution
of the $x_n$ given by
equation (\ref{eq}) with $\beta = \beta'$. This situation
would apply to a very mobile tracer diffusing in a background of very
slowly moving charges of the same valence, for example an electron in
a background of immobile ions. To facilitate the calculation of
$\overline{\kappa}$ we pass to the grand canonical ensemble for the
scatters to obtain
\begin{eqnarray}
\fl
\overline{\kappa} = &{1\over \Xi}&\sum_{N_+,N_-}\int {d^Dx_i\over N_+! N_-
!}{d^Dx\over V} \nonumber \\
\fl & &z^{N_+ + N_-}\exp\left(-\beta\sum_{i<j}q_i q_j
V(x_i-x_j)\right)\exp\left( \beta \sum_{i}q_iV(x_i-x)\right)~~,
\end{eqnarray}
where the integration over $x$ above is the spatial averaging of
$\kappa(x)$ over $V$, $\Xi$ is the grand partition function and $z$ is
the fugacity of the positive and negative charges.  The above
expression can clearly be written as
\begin{equation}
{\overline{\kappa}} = {1\over z V \Xi} \sum_{N_+,N_-} {(N_-+1)} z^{N_+
+N_-+1}Z_{N_+,N_-+1}~~,
\end{equation}
where $Z_{N_+,N_-}$ is the canonical partition function for a system
of $N_{+/-}$ positive/negative charges. In the thermodynamic limit we
thus obtain
\begin{equation}
{\overline{\kappa}} = {\rho \over 2 z}~~,
\end{equation}
which yields
\begin{equation}
{\kappa_e^{(g)}\over \kappa_0} = {2 z\over \rho}~~.
\end{equation}

For the case of a pure Coulomb interaction {\em i.e.} when $\mu = 0$
\begin{equation}
V(x)= -\ln({ \vert x\vert\over L})~~,
\end{equation}
where $L$ is a length scale which from here on we set to be $1$.  The
statistical mechanics of the two dimensional Coulomb gas has been
recently exactly solved \cite{sata} in the region where the model is
thermodynamically stable ({\em i.e.} $\beta <2$). For $\beta >2$ the
system is thermodynamically unstable and collapses unless a hard core
interaction is included between the particles; this instability is
present for the reasons mentioned previously. In the stable region it
has been shown
\begin{equation}
{\rho^{1-\beta/4}\over z} = 2 \left(\pi \beta\over
8\right)^{\beta/4}{\Gamma(1 -\beta/4)\over \Gamma(1 +
\beta/4)}\left[_2 F_1\left({1\over 2}, {\beta\over 4-\beta}; 1 +
{\beta\over 2(4-\beta)};1\right)\right]^{1-\beta/4}~~.
\end{equation}
The resulting behaviour for $\kappa_e^{(g)}$ when $\beta \ll1$ is
\begin{equation}
{\kappa_e^{(g)}\over \kappa_0} \approx
(\rho\beta)^{-\beta/4}\exp\left[ -\left(2\gamma +
\ln({\pi\over2})\right) {\beta\over 4} - {7\over
6}\zeta(3)\left({\beta\over 4}\right)^3 -\zeta(3)\left({\beta\over
4}\right)^4 \right]~~,
\end{equation}
where $_2F_1$ the the hyper-geometric function.
Near the collapse point one finds
\begin{equation}
{\kappa_e^{(g)}\over \kappa_0} \approx \left({2 -\beta\over
\rho\pi}\right)^{1/2} \label{vdc}
\end{equation}
as $\beta \to 2^{-}$.  The interesting thing about these results is
that the diffusivity is non-zero in all the region of stability, in
contrast to the case of uniformly distributed Coulomb scatterers where
it is zero in the thermodynamic limit in this region.  Physically the
fact that the Coulomb scatterers are allowed to equilibrate generates
a screening characterised by a Debye length and the long range
fluctuations of the electrostatic field are thus suppressed allowing a
normal diffusion. The way in which the diffusion constant vanishes in
equation (\ref{vdc}) as $\beta \to 2^-$ is also clearly very different
to the case of a uniform distribution of Yukawa scatterers.
\section{Conclusion}
We have examined the problem of calculating the effective diffusivity
of a Langevin particle diffusing in the force field generated by
charged scatterers in two dimensions. The fact that the system is
statistically electro-neutral means that the potential $\phi$
has the same distribution as $-\phi$. Using a general exact result
connecting the effective diffusions constant of the gradient flow 
and varying permeability problem and an exact result in two dimensions
for the permeability problem allows us to solve exactly the problem 
of calculating $\kappa_e^{(g)}$ in the cases studied here. 

The results obtained give insight into the  circumstances under which the
random field may be taken to be Gaussian for the purposes of calculating
the effective diffusion constant. It is clear from this study that 
the random field due  to frozen scatterers will lead to transport 
properties that are generally different to that of a Gaussian random 
field with the same two point correlation function. 
It was also shown that the nature
of the correlations between the point scatterers can drastically 
modify the behaviour of  $\kappa_e^{(g)}$. For instance the diffusion
constant for a system electro-neutral  uniformly distributed Coulomb
scatterers is always zero at finite temperature, however if they
are taken to in an equilibrium configuration of the same temperature
as the tracer particles then  $\kappa_e^{(g)}$ is finite at high
temperature.  The exact results here should be useful in developing
perturbative techniques to treat  problems in higher dimensions and
in two dimensions where there is only one type of scattering particle, 
{\em i.e.} no charge.

\end{document}